\documentclass[a4paper]{article}
\usepackage{adjustbox}
\usepackage{graphicx}
\usepackage{multirow}
\usepackage[table,xcdraw]{xcolor}
\graphicspath{ {./images/} }
\usepackage[numbers,sort&compress]{natbib}
\usepackage{INTERSPEECH2020}
\usepackage{url}
\usepackage[colorlinks,citecolor=blue,urlcolor=blue,pdfsubject={LaTeX}]{hyperref}
\PassOptionsToPackage{hyphens}{url}
\usepackage[hyphenbreaks]{breakurl}
\usepackage[keeplastbox]{flushend}

\title{An ASR Guided Speech Intelligibility Measure for TTS Model Selection}
\name{Arun Baby, Saranya Vinnaitherthan, Nagaraj Adiga, Pranav Jawale,\\Sumukh Badam, Sharath Adavanne, Srikanth Konjeti}
\address{Zapr Media Labs (Red Brick Lane Marketing Solutions Pvt. Ltd.), India}
\email{arun.baby@zapr.in}

\begin{document}

\maketitle
\begin{abstract}
The perceptual quality of neural text-to-speech (TTS) is highly dependent on the choice of the model during training. Selecting the model using a training-objective metric such as the least mean squared error does not always correlate with human perception. In this paper, we propose an objective metric based on the phone error rate (PER) to select the TTS model with the best speech intelligibility. The PER is computed between the input text to the TTS model, and the text decoded from the synthesized speech using an automatic speech recognition (ASR) model, which is trained on the same data as the TTS model. With the help of subjective studies, we show that the TTS model chosen with the least PER on validation split has significantly higher speech intelligibility compared to the model with the least training-objective metric loss. Finally, using the proposed PER and subjective evaluation, we show that the choice of best TTS model depends on the genre of the target domain text. All our experiments are conducted on a Hindi language dataset. However, the proposed model selection method is language independent.

\end{abstract}

\section{Introduction}
The recent advancement in deep learning techniques has enabled end-to-end text-to-speech (TTS) systems to achieve perceptual quality comparable to human speech~\cite{oord2016wavenet, shen2018natural, ren2019fastspeech, kons2019high}. Usually, such TTS systems are trained for a large number of epochs from which a final model is selected for production. One of the approaches for selection is to choose the model with the least training-objective loss on an unseen validation split, for example, the mean squared error (MSE) computed on mel spectrogram representation~\cite{shen2018natural}. However, the computation of this loss requires both text sequences and corresponding audio recordings that may not always available. More importantly, the TTS model chosen with the least training-objective loss does not always correlate to the best perceptual quality~\cite{Wang2017, theis2015note}.

One of the approaches to choose a TTS model with the best perceptual quality is to manually rate a few synthesized recordings for different TTS models and choose the best model. These TTS models for manual listening tests are sampled randomly from the region where the training-objective loss on the training split has saturated. Mean opinion score (MOS) is a popular choice for such perceptual quality rating and is most often used to evaluate either the naturalness or the intelligibility of the synthesized speech. Naturalness refers to how close the synthesized speech is to real-life speech, and intelligibility refers to the ease of understanding the spoken content. However, performing such a subjective analysis of multiple recordings across models can be time-consuming, expensive, and not scalable.

To overcome the scalability issue of MOS, several objective metrics have been proposed to replicate the MOS results. Specifically, these metrics propose to replace the MOS of speech intelligibility using modules from automatic speech recognition (ASR)~\cite{vich2008automatic, cervnak2009diagnostic, wang2012objective, ullmann2015withReference, ullmann2015withoutReference}. Cer\v{n}ak et. al.~\cite{cervnak2009diagnostic} employed subjective analysis and showed that the word error rate (WER) computed between the input text to TTS, and decoded text of synthesized speech using an ASR system correlates with speech intelligibility. In~\cite{ullmann2015withReference}, a phonetic acoustic model was employed to estimate the phone posterior probability sequence of both the reference and synthesized recordings. The distance between them was computed using a dynamic time warping algorithm. This distance was observed to correlate with subjective speech intelligibility. However, this metric requires the reference audio recording, which may not always be available. In~\cite{lo2019mosnet}, a deep learning-based model was trained to directly estimate the MOS for the synthesized audio. This model was trained on a dataset of synthesized speech and their corresponding MOS. However, the collection of such datasets for different languages and objectives of naturalness and intelligibility is not trivial. All of the existing works have only studied the correlation between speech intelligibility and their proposed objective metrics for their final chosen TTS model. However, to the best of the authors' knowledge, there have been no studies on the selection of the final TTS model itself that has the best perceptual quality, such as speech intelligibility.

In this paper, we propose to use ASR based phone error rate (PER) as the objective metric for choosing the TTS model with the best speech intelligibility. The PER is computed between the input text to the TTS system and the decoded text of the TTS synthesized speech using an ASR system. We propose to employ PER for model selection during TTS system training. Using subjective preference tests, we have validated that the PER chosen model has better speech intelligibility compared to a model chosen using the least training-objective loss on validation split. A similar character error rate metric is employed in the popular TTS framework of ESPnet~\cite{hayashi2019espnettts}. However, they only use it to automatically detect the alignment failures in the synthesized speech and not as an objective metric to evaluate the speech intelligibility. 

Finally, we show that the choice of the TTS model depends on the genre of the target domain text. We study this by comparing the model selection using PER for two separate validation splits - the first with similar text as the training split, and the second with text from a different genre (news) with 57.1\% unseen unique words. The proposed model selection criteria chose different epochs for the two validation splits, that each corresponded to the best speech intelligibility for the respective splits according to the subjective analysis. 

In this paper, we only study the model selection for neural TTS frontend of Tacotron 2~\cite{shen2018natural}, while keeping the vocoder fixed across our experiments. The proposed method itself is generic and can be employed for other popular TTS frontends.

\begin{figure}[!htp]
\centerline{\includegraphics[width=\linewidth]{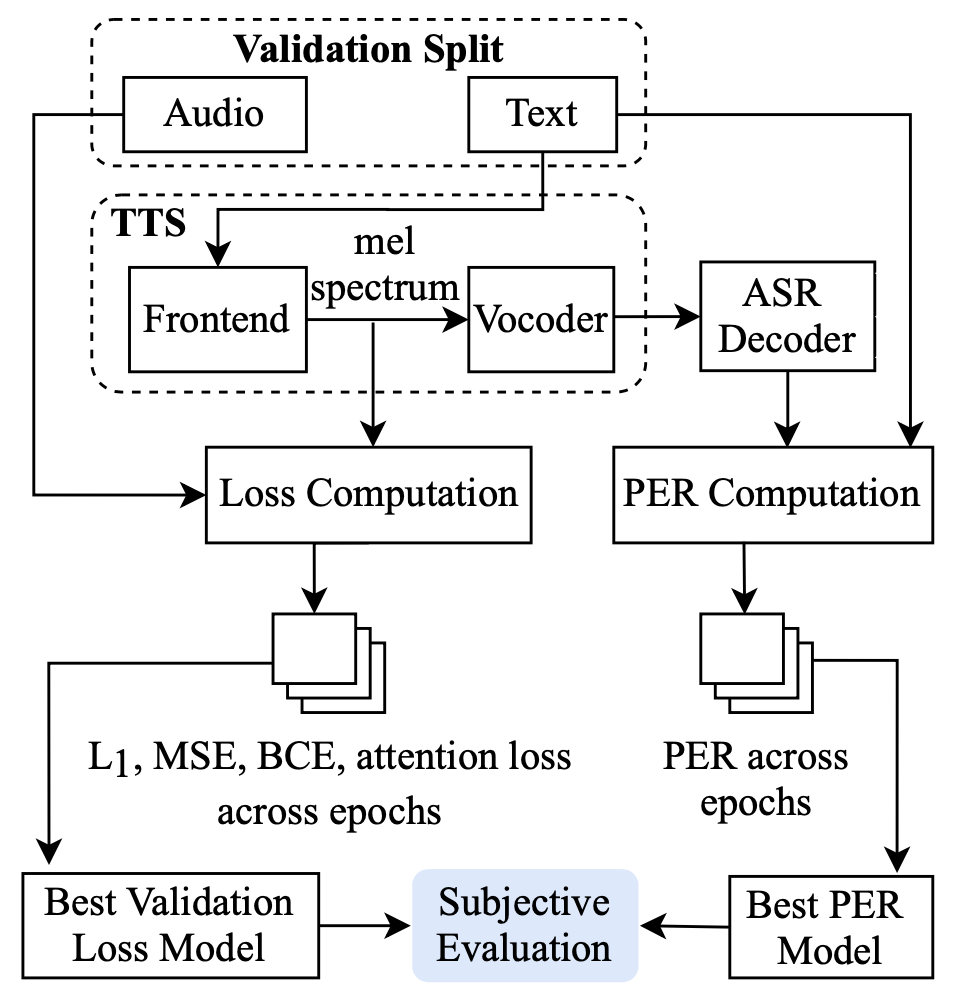}}
\caption{Block diagram of TTS model selection using PER.}
\label{pereval}
\end{figure}

\section{Proposed method}
The block diagram of the proposed model selection method is shown in Figure~\ref{pereval}. The TTS module consists of a frontend and a vocoder. Given an input text, the frontend maps it to a sequence of mel spectrogram features corresponding to a spoken version of the input text. These features are further mapped to an audio waveform using a vocoder. In this paper, we only study the selection of the frontend model and use a fixed vocoder across experiments. Hereafter, model selection refers only to the frontend model selection.

In this paper, we propose to employ PER as the objective metric for speech intelligibility and choose the model from the training epoch that has the least PER on validation split. To compute PER, we first decode the phone sequence from the vocoder output using an ASR system. The PER is then computed using the following equation between the decoded phone sequence and the corresponding input text to the TTS module, at the phone level.
\begin{equation}
    PER = \frac{S + D + I}{N},
\end{equation}
where \textit{S}, \textit{D} and \textit{I} represents the number of substitutions, deletions and insertions, respectively, between the decoded and the reference phone sequence. \textit{N} represents the number of phones in the reference text. Finally, we compare the speech intelligibility of the best PER model with the best training-objective loss model using subjective evaluation (described in Section~\ref{sec:exp}).

We propose to use PER instead of the popular WER as it provides more insights about the TTS model itself. For example, from the analysis of the PER results, the observation of a particular phone being deleted often throws light on poor modeling of the phone, potentially due to lack of sufficient training examples. Similarly, the observation of a phone being substituted often with another phone can be a result of confusing pronunciation by the speaker in the training data. Such insights cannot be obtained directly using WER as an objective metric.

The TTS and ASR systems are both trained using identical training and validation splits from the same dataset. The details of these individual TTS and ASR systems are discussed below.

\subsection{TTS system}
\label{ttssystem}
A neural TTS system is generally comprised of a frontend and vocoder as shown in Figure~\ref{pereval}. As the fronted, we use the Tacotron 2 (v3) recipe of ESPnet~\cite{hayashi2019espnettts}, which is an auto-regressive based sequence-to-sequence model with a location-sensitive and guided attention mechanism. The frontend is trained with the phone sequence of the input text, and the 80-band mel spectrogram feature of the corresponding audio, computed with a 1024 point discrete Fourier transform and 256 sample hop-length. The frontend is trained for 800 epochs with a batch size of 56 on 4 GPUs. 

During each epoch of the training, we compute the training-objective loss on the validation split. The training-objective loss in the Tacotron 2 (v3) recipe is the sum of the MSE, $l_1$, binary cross-entropy (BCE), and attention loss. Where the MSE and $l_1$ are computed between the predicted and reference mel spectrograms, BCE is computed for stop-token prediction, and the attention loss measures the diagonality of alignment between the encoder and decoder outputs. Finally, the model during the epoch with the least training-objective loss on the validation split is chosen as the best validation loss model. 

The mel spectrogram output of fronted is mapped to waveform using the parallel wavegan (PWG) vocoder~\cite{yamamoto2020parallel}. The PWG vocoder is a non-autoregressive variant of the WaveNet~\cite{oord2016wavenet} vocoder that has a significantly faster inference time. We use the publicly available implementation of PWG, whose code is accessible here\footnote{https://github.com/kan-bayashi/ParallelWaveGAN}. The ASR system discussed in the next section is then used to decode the phone sequence from the PWG synthesized audio. Since the PER results are only meaningful after the initial convergence of the frontend model, we start computing the PER after epoch 50 and thereafter compute it for every 10 epochs. Finally, we choose the model with the least PER on validation split as the best PER model. 

\subsection{ASR system}
\label{asrsystem}
The ASR system is built using the Kaldi toolkit~\cite{povey2011kaldi}. We initially train the acoustic model using the Gaussian mixture model-hidden Markov model (GMM-HMM) recipe\footnote{wsj/s5/run.sh} and obtain alignments from the triphone model. Thereafter, these alignments are used to train our ASR system using the deep neural network-hidden Markov model (DNN-HMM) recipe\footnote{wsj/s5/local/chain/tuning/run\_tdnn\_1g.sh}. The default augmentations of speed and volume perturbations, that are part of the DNN-HMM recipe are used during training.

The ASR system uses phone-based language models (LM) built using the SRILM toolkit~\cite{stolcke2002srilm} during decoding.

\begin{figure}[!tp]
    \centering
    \centerline{\includegraphics[width=\linewidth]{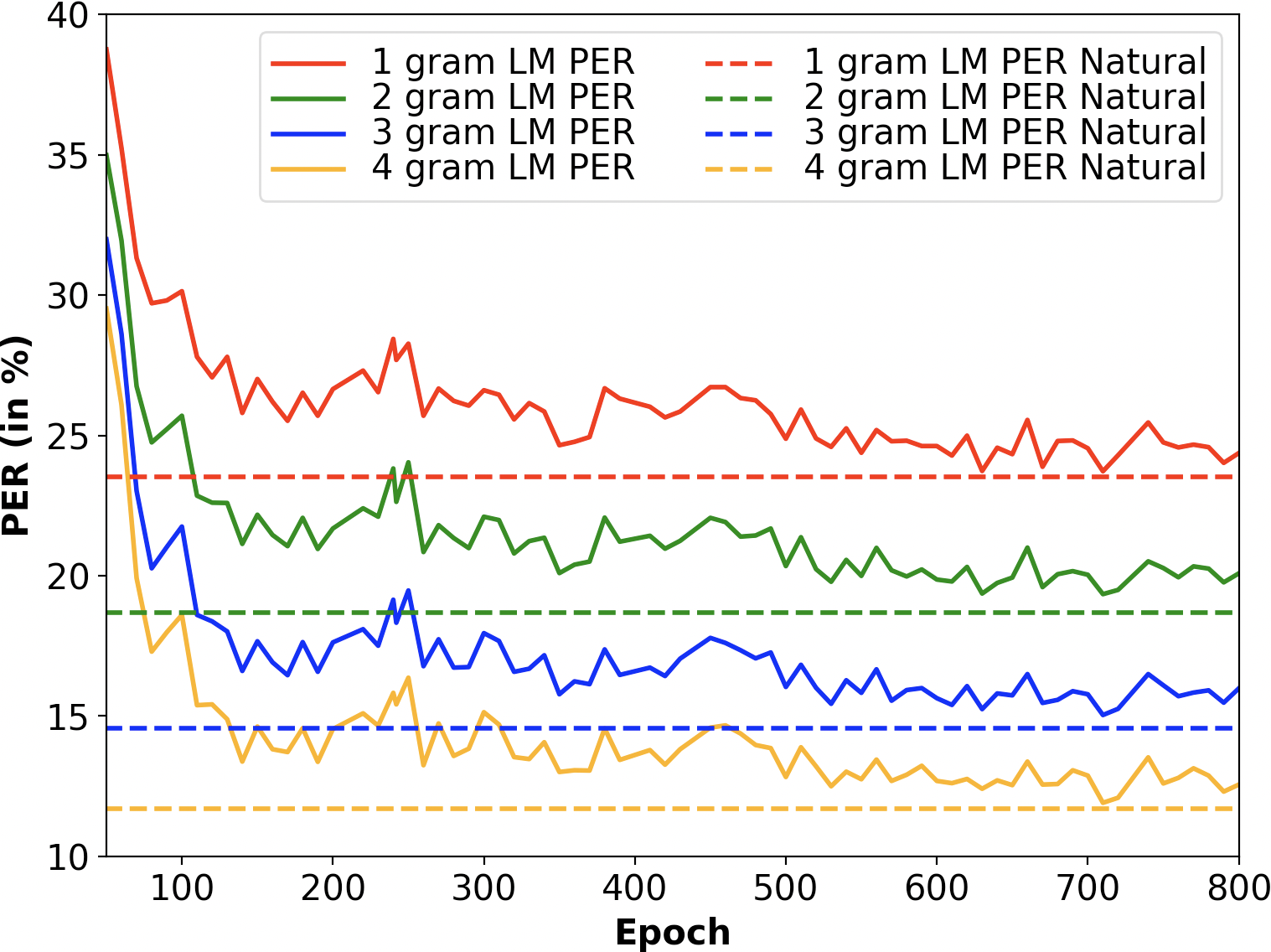}}
    \caption{PER across training epochs for synthesized (solid lines) and original (dashed lines) recordings using different language models (LM) in an ASR system.}
    \label{per_all}
\end{figure}

\section{Evaluation}
\subsection{Dataset}
As the dataset, we use a subset of IndicTTS~\cite{babycbblr2016} dataset of Hindi language and male speaker. This dataset consists of about 9 hours of studio-quality recording by a professional voice over artist. The original recordings are at a 48 kHz sampling rate, however, all the studies in this paper are performed at 16 kHz. The genre of the spoken text is fiction and children's stories. We map this to phonemic text with Indian common label set phones~\cite{ramani2013common} using a grapheme to phoneme model~\cite{nlp:tsd16conf, arunThesis}.  

Among the 6032 sentences in the dataset, we randomly choose 5732 sentences (8.76 hours) as the \textit{training split} and the remaining 300 sentences (0.44 hours) as the \textit{validation split}. The text in the training split has about 13,536 unique words. Both the ASR and TTS modules are trained using these training and validation splits using phonetic transcripts of 60 unique phones. The TTS modules are additionally trained with five punctuations and a word boundary. Similarly, the ASR system consists of an additional silence-phone.  All our objective and subjective studies regarding the synthesis quality are carried out on the \textit{validation split}, which we also refer to as the \textit{in-domain test data}. This data has about 1887 unique words, of which 21\% words are unseen, and it only consists of 55 of the 60 total phones in the training split.

To study the synthesis quality on the text of a different genre, we manually curated an \textit{out-of-domain test data}, that has no audio as follows. In contrast to the training split that mainly consists of story genre text, we scraped news genre text from online resources. From which we chose sentences that had a word length in the range of 3-6. This test data consists of 1239 unique words of which 57.1\% are unseen, and it only consists of 50 of the 60 total phones in the training split.

\subsection{Experiments}
\label{sec:exp}

To verify the role of LM in the proposed model selection approach, we first study the choice of LM for ASR system. We generate 1-, 2-, 3- and 4-gram LMs on the training split using SRILM and study the model selection on in-domain test data.

The choice of PER as a speech intelligibility metric is evaluated with subjective hearing tests. The best models for the in-domain test data, based on the validation loss and PER are used to synthesize the text of in-domain test data. A pair-wise comparison test is conducted between the recordings of the two best models with both native (L1) and non-native (L2) speakers. During the test, the participants were given 30 pairs of synthesized recordings randomly chosen from the 300 recordings of in-domain test data. The participants were asked to rate the speech intelligibility by choosing either of the two recordings or both if they were equally intelligible. There were 20 participants in total, with equal distribution of L1 and L2 speakers.

Finally, we study the model selection for the text of a different genre from training data. The best PER model is selected based on out-of-domain test data whose genre is different from the training split. The speech intelligibility of the best out-of-domain test data model is then compared using subjective pair-wise evaluations with both the best validation loss model and the best in-domain test data model.

The samples used for the subjective evaluations in our paper are available at \url{https://www.zapr.in/interspeech2020/samples}.

\begin{figure}[!tp]
    \centering
    \includegraphics[width=\linewidth]{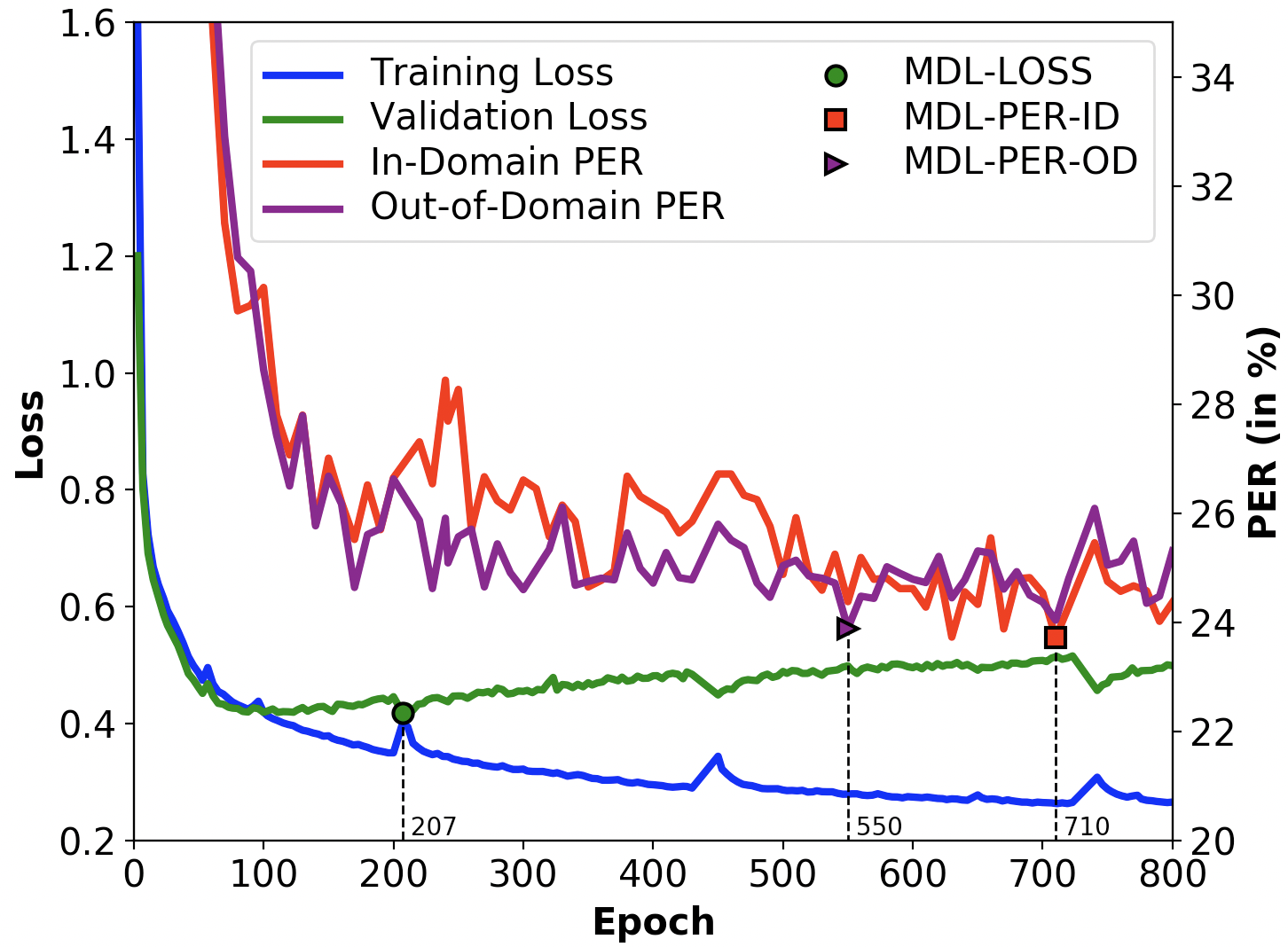}
    \caption{The training and validation loss curves, in-domain and out-of-domain PERs across epochs. The best validation loss model MDL-LOSS is chosen for the epoch with the least validation loss. Similarly, the best PER models of in-domain MDL-PER-ID and out-of-domain MDL-PER-OD are chosen for the corresponding epochs with the least PER.}
    \label{mse_per}
\end{figure}

\section{Results and discussion}
\subsection{Language model selection}
The PER values for synthesized recordings of in-domain test data obtained using different LMs are shown as solid lines in Figure~\ref{per_all}. The dotted lines with similar color as solid lines represent the corresponding PER scores for the original recordings of in-domain test data. From the figure, it is clear that irrespective of the chosen LM, the PER curve trends are similar and the corresponding epoch with minimum PER are identical across the curves. Hence, hereafter we only use the 1-gram LM for our studies. The choice of 1-gram additionally removes the dependency of ASR output on neighboring phones. This enables a more reliable understanding of individual phone modeling in TTS frontend based on the PER results.

\begin{table}[]

\begin{adjustbox}{max width=8cm}
\centering

\begin{tabular}{|c|c|c|c|c|}
\hline
{\color[HTML]{000000} }                                  & {\color[HTML]{000000} }                                                                                      & \multicolumn{2}{c|}{{\color[HTML]{000000} \textbf{Correctly detected}}}             & {\color[HTML]{000000} }                                                                                       \\ \cline{3-4}
\multirow{-2}{*}{{\color[HTML]{000000} \textbf{Phones}}} & \multirow{-2}{*}{{\color[HTML]{000000} \textbf{\begin{tabular}[c]{@{}c@{}}Total\\ Occurences\end{tabular}}}} & {\color[HTML]{000000} \textbf{MDL-LOSS}} & {\color[HTML]{000000} \textbf{MDL-PER-ID}} & \multirow{-2}{*}{{\color[HTML]{000000} \textbf{\begin{tabular}[c]{@{}c@{}}Difference \\ count\end{tabular}}}} \\ \hline \hline
{\color[HTML]{000000} \textbf{/a/}}                        & {\color[HTML]{000000} 1784}                                                                                  & {\color[HTML]{000000} 1395}              & {\color[HTML]{000000} \textbf{1460}}       & {\color[HTML]{000000} 65}                                                                                     \\ \hline
{\color[HTML]{000000} \textbf{/ee/}}                       & {\color[HTML]{000000} 1407}                                                                                  & {\color[HTML]{000000} 1203}              & {\color[HTML]{000000} \textbf{1261}}       & {\color[HTML]{000000} 58}                                                                                     \\ \hline
{\color[HTML]{000000} \textbf{/i/}}                        & {\color[HTML]{000000} 717}                                                                                   & {\color[HTML]{000000} 464}               & {\color[HTML]{000000} \textbf{498}}        & {\color[HTML]{000000} 34}                                                                                     \\ \hline
{\color[HTML]{000000} \textbf{/k/}}                        & {\color[HTML]{000000} 1217}                                                                                  & {\color[HTML]{000000} 1061}              & {\color[HTML]{000000} \textbf{1095}}       & {\color[HTML]{000000} 34}                                                                                     \\ \hline
{\color[HTML]{000000} \textbf{/h/}}                        & {\color[HTML]{000000} 676}                                                                                   & {\color[HTML]{000000} 497}               & {\color[HTML]{000000} \textbf{528}}        & {\color[HTML]{000000} 31}                                                                                     \\ \hline\hline
{\color[HTML]{000000} \textbf{/z/}}                        & {\color[HTML]{000000} 55}                                                                                    & {\color[HTML]{000000} \textbf{42}}       & {\color[HTML]{000000} 40}                  & {\color[HTML]{000000} -2}                                                                                     \\ \hline
{\color[HTML]{000000} \textbf{/uu/}}                       & {\color[HTML]{000000} 131}                                                                                   & {\color[HTML]{000000} \textbf{97}}       & {\color[HTML]{000000} 94}                  & {\color[HTML]{000000} -3}                                                                                     \\ \hline
{\color[HTML]{000000} \textbf{/gh/}}                       & {\color[HTML]{000000} 33}                                                                                    & {\color[HTML]{000000} \textbf{31}}       & {\color[HTML]{000000} 27}                  & {\color[HTML]{000000} -4}                                                                                     \\ \hline
{\color[HTML]{000000} \textbf{/sh/}}                       & {\color[HTML]{000000} 123}                                                                                   & {\color[HTML]{000000} \textbf{99}}       & {\color[HTML]{000000} 95}                  & {\color[HTML]{000000} -4}                                                                                     \\ \hline
{\color[HTML]{000000} \textbf{/q/}}                        & {\color[HTML]{000000} 633}                                                                                   & {\color[HTML]{000000} \textbf{145}}      & {\color[HTML]{000000} 137}                 & {\color[HTML]{000000} -8}                                                                                     \\ \hline
\end{tabular}
\end{adjustbox}
\vspace{10pt}
\caption{The number of phone instances that were decoded correctly using an ASR system for recordings synthesized using MDL-LOSS and MDL-PER-ID models. The table only presents the top and bottom five phones with difference in phone detection between the two models. }
\label{phoneme_detect}
\vspace{-10pt}
\end{table}

\subsection{In-domain analysis}

The training-objective loss on the training split, the validation (in-domain test data) split, and the corresponding PER across different epochs of training is visualized in Figure~\ref{mse_per}. The best model selected based on the validation loss curve is at epoch 207, hereafter referred to as MDL-LOSS. However, according to the proposed metric PER, the model with the best speech intelligibility was at epoch 710, hereafter referred to as MDL-PER-ID. Figure~\ref{per_all} gives a different perspective for the same choice of epoch 710, where the PER for synthesized recordings (solid lines) converges to the PER with the original in-domain test data recordings (dashed lines). Further, although both the PER and validation losses are computed on the same in-domain test data, according to the validation loss curve in Figure~\ref{mse_per}, the model started over-fitting after epoch 207. However, according to the PER curve, no such over-fitting is observed, and the model continues to improve the speech intelligibility until the end of the training (800 epochs).

To study the improvement in speech intelligibility between the two models - MDL-LOSS and MDL-PER-ID, we analyzed the phone detection statistics. The ASR system was used to decode the phone sequences for all the recordings synthesized with the in-domain text, using both the models. The correct phones detection according to PER was studied. The original test data consists of 55 unique phones with a total occurrence of 18251. Among which, the MDL-PER-ID correctly identified 13983 instances compared to the 13612 instances by MDL-LOSS. This improvement in the detection of phones by MDL-PER-ID was a result of improved detection of 29 of the 55 unique phones, which we correlate to better speech intelligibility in synthesis. The detection of 15 other phones remained unchanged and the performance reduced marginally for the rest. Table~\ref{phoneme_detect} shows the top five phones whose detection improved and the bottom five phones whose detection reduced. We see that the overall difference in improved detection is much larger compared to the phones whose detection reduced.

The results of the subjective analysis for the in-domain recordings synthesized by MDL-LOSS and MDL-PER-ID is shown in Table~\ref{pctest}. Overall, 42.8\% of participants chose the proposed MDL-PER-ID recordings, in comparison to 28.1\% who preferred MDL-LOSS recordings. Both the L1 and L2 speakers preferred the proposed MDL-PER-ID recordings, and this preference was stronger for L2 speakers. These results prove that the PER can be used as an objective metric for speech intelligibility. Correspondingly, the TTS model chosen with the best PER on the validation data correlates to synthesis with better speech intelligibility, compared to a model chosen based on training-objective loss.

\subsection{Out-of-domain analysis}

The PER curve for the out-of-domain test data is shown in Figure~\ref{mse_per}. In comparison to the best in-domain test data model MDL-PER-ID, which was chosen at epoch 710, the model with the best speech intelligibility for the out-of-domain test data MDL-PER-OD is obtained at epoch 550. This shows that the best PER model is dependent on the target domain genre of the text. In the phone detection studies, among the 9763 total phones the MDL-PER-OD detected 7468 of them correctly, compared to 7345 from MDL-LOSS. This was a result of improved detection of 29 of the 50 unique phones in comparison to MDL-LOSS. The detection of 9 other phones remained unchanged, and the rest reduced marginally.

We performed two separate pair-wise subjective studies on MDL-PER-OD. The first (\emph{Exp.\,1}), between the best validation loss model MDL-LOSS and MDL-PER-OD. The results of which are presented in Table~\ref{pctest}. Similar to the results of in-domain analysis, both L1 and L2 participants continued to prefer the proposed PER based MDL-PER-OD recordings. Further, in comparison to the MDL-LOSS recordings, the preference for MDL-PER-OD recordings in the out-of-domain analysis was much stronger than the preference for MDL-PER-ID recordings in the in-domain analysis. We conducted a second experiment (\emph{Exp.\,2}) on the out-of-domain text to check if the participants would continue to prefer MDL-PER-OD recordings over MDL-PER-ID recordings. From the Table~\ref{pctest}, we observe that both L1 and L2 participants continued to prefer MDL-PER-OD recordings for out-of-domain text. However, the preference of L1 speakers for MDL-PER-OD recordings was not as strong as the L2 speakers. Further, nearly 50\% of the participants, felt both the models were equally intelligible. In general, these results show that when the genre of the target domain text is different from the training data, the PER based model selection is better than using a generic validation loss based model across different target text genres.

\begin{table}[]
\centering

\begin{adjustbox}{max width=8cm}
\begin{tabular}{|c|c|c|c|c|}
\hline
{\color[HTML]{000000} }                                                                                            & {\color[HTML]{000000} \textbf{Speakers}}   & {\color[HTML]{000000} \textbf{MDL-LOSS}}                                              & {\color[HTML]{000000} \textbf{\begin{tabular}[c]{@{}c@{}}MDL-\\ PER-ID\end{tabular}}}  & {\color[HTML]{000000} \textbf{Both}} \\ \cline{2-5} 
{\color[HTML]{000000} }                                                                                            & {\color[HTML]{000000} L1}               & {\color[HTML]{000000} 27.5}                                                           & {\color[HTML]{000000} 40.0}                                                              & {\color[HTML]{000000} 32.5}             \\ \cline{2-5} 
{\color[HTML]{000000} }                                                                                            & {\color[HTML]{000000} L2}               & {\color[HTML]{000000} 28.8}                                                           & {\color[HTML]{000000} 45.6}                                                            & {\color[HTML]{000000} 25.6}             \\ \cline{2-5} 
\multirow{-4}{*}{{\color[HTML]{000000} \begin{tabular}[c]{@{}c@{}}In-domain\\ Analysis\end{tabular}}}              & {\color[HTML]{000000} \textbf{Overall}} & {\color[HTML]{000000} 28.1}                                                           & {\color[HTML]{000000} 42.8}                                                            & {\color[HTML]{000000} 29.1}             \\ \hline \hline
{\color[HTML]{000000} }                                                                                            & {\color[HTML]{000000} \textbf{Speakers}}   & {\color[HTML]{000000} \textbf{MDL-LOSS}}                                              & {\color[HTML]{000000} \textbf{\begin{tabular}[c]{@{}c@{}}MDL-\\ PER-OD\end{tabular}}} & {\color[HTML]{000000} \textbf{Both}} \\ \cline{2-5} 
{\color[HTML]{000000} }                                                                                            & {\color[HTML]{000000} L1}               & {\color[HTML]{000000} 30.0}                                                             & {\color[HTML]{000000} 44.4}                                                            & {\color[HTML]{000000} 25.6}             \\ \cline{2-5} 
{\color[HTML]{000000} }                                                                                            & {\color[HTML]{000000} L2}               & {\color[HTML]{000000} 15.0}                                                             & {\color[HTML]{000000} 53.3}                                                            & {\color[HTML]{000000} 31.7}             \\ \cline{2-5} 
\multirow{-4}{*}{{\color[HTML]{000000} \begin{tabular}[c]{@{}c@{}}Out-of-domain\\ Analysis\\ \emph{Exp.\,1}\end{tabular}}} & {\color[HTML]{000000} \textbf{Overall}} & {\color[HTML]{000000} 22.5}                                                           & {\color[HTML]{000000} 48.9}                                                            & {\color[HTML]{000000} 28.6}               \\ \hline \hline
{\color[HTML]{000000} }                                                                                            & {\color[HTML]{000000} \textbf{Speakers}}   & {\color[HTML]{000000} \textbf{\begin{tabular}[c]{@{}c@{}}MDL-\\ PER-ID\end{tabular}}} & {\color[HTML]{000000} \textbf{\begin{tabular}[c]{@{}c@{}}MDL-\\ PER-OD\end{tabular}}} & {\color[HTML]{000000} \textbf{Both}} \\ \cline{2-5} 
{\color[HTML]{000000} }                                                                                            & {\color[HTML]{000000} L1}               & {\color[HTML]{000000} 33.3}                                                               & {\color[HTML]{000000} 34.0}                                                                & {\color[HTML]{000000} 32.7}                 \\ \cline{2-5} 
{\color[HTML]{000000} }                                                                                            & {\color[HTML]{000000} L2}               & {\color[HTML]{000000} 12.7}                                                               & {\color[HTML]{000000} 20.0}                                                                & {\color[HTML]{000000} 67.3}                 \\ \cline{2-5} 
\multirow{-4}{*}{{\color[HTML]{000000} \begin{tabular}[c]{@{}c@{}}Out-of-domain\\ Analysis\\ \emph{Exp.\,2}\end{tabular}}} & {\color[HTML]{000000} \textbf{Overall}} & {\color[HTML]{000000} 23.0}                                                          & {\color[HTML]{000000} 27.0}                                                           & {\color[HTML]{000000} 50.0}            \\ \hline
\end{tabular}
\end{adjustbox}
\vspace{10pt}
\caption{Pair-wise comparison results for subjective evaluation of speech intelligibility (values in percentages).}
\label{pctest}
\vspace{-10pt}
\end{table}

\section{Conclusion}
In this work, we proposed to employ PER as an objective metric to select the TTS model with the best speech intelligibility. Our preference test studies show that the TTS model with the best PER on validation split has better speech intelligibility compared to the model chosen based on minimum training-objective loss on validation split. Finally, using the PER metric and subjective preference test on unseen out-of-domain text, we show that the choice of the best TTS model is dependent on the target application text genre. All the experiments were conducted on a Hindi language dataset. However, the proposed method itself is language independent. 


\bibliographystyle{IEEEtran}

\bibliography{main}

\end{document}